\documentstyle[12pt,aaspp4]{article}
\begin{document}
\title{Discovery of 5 Minutes Sine-like Oscillations in the light curve
of the Asteroid 1689 Floris-Jan}
\author{W. Pych}
\affil{Warsaw University Observatory, Al. Ujazdowskie 4, 00-478 Warszawa,
Poland}
\affil{\tt e-mail: pych@sirius.astrouw.edu.pl}
\begin{abstract}
We present CCD photometry of the long period asteroid 
1689 Floris-Jan. On the light curve from nights 1997.02.10/11 and
1997.02.11/12 we detected sine-like oscillations with the period
P=4.98 ${\pm}$ 0.01 minutes and full amplitude about 0.11 mag.
Observations from night 1997.03.07/08 show no light variations at
this period.
\end{abstract}
%
%
\section{Introduction}
	The asteroid 1689 Floris-Jan is a small sized object from the
main asteroid belt. It was discovered on September 1930 by H. Van Gent.
The diameter of 1689 Floris-Jan is estimated to be 9 to 27 km
depending on the assumed albedo. In 1982 Schober et. al. (1982) presented UBV
photometry of this object, which revealed its unusually long period of
rotation of 6.042 ${\pm}$ 0.21 days. Typical periods
of rotation of asteroids range from about 2 to about 20 hours. For many years
this asteroid was considered to be one of the slowest rotators among minor
planets. 1689 Floris-Jan was selected as a target for the observations by
our colleagues from the Astronomical Observatory of Adam Mickiewicz University,
Pozna{\'n}, during a long term program of CCD photometry of minor
planets (Kryszczy{\'n}ska et. al., 1996).
\section{Observations, data reduction and photometry}
	Observational data were collected at the Ostrowik Station of the
Warsaw University Observatory. We used a CCD system attached to 0.6 m
telescope (Udalski and Pych, 1992). On nights of February 10th, 1997
and February 11th, 1997 we used Cousins I filter for 120 second exposures.
On the night of March 7th, 1997 we made unfiltered
exposures with integration time of 60 seconds.
	De-biasing and flat-fielding were done with the procedures
available in the IRAF\footnote{IRAF is distributed by the National
Optical Astronomical
Observatories, operated by the Association of Universities for Research
in Astronomy, Inc., under contact with the National Science Foundation}
 package. Aperture photometry was obtained with the
DaophotII program (Stetson, 1987).
\section{Results and data analysis}
	Figure 1 presents the light curve of the 1689 Floris-Jan compared
with the light curves of two nearby stars, obtained on the night
1997.02.11/12. Monotonic rise of brightness is observed. This is the
manifestation of the long rotational period of the planet.
Short period light variations with an amplitude of about 0.1 mag. are
superimposed onto the long term trend.
\begin{figure}[htbp]
\vspace{9cm}
\includegraphics{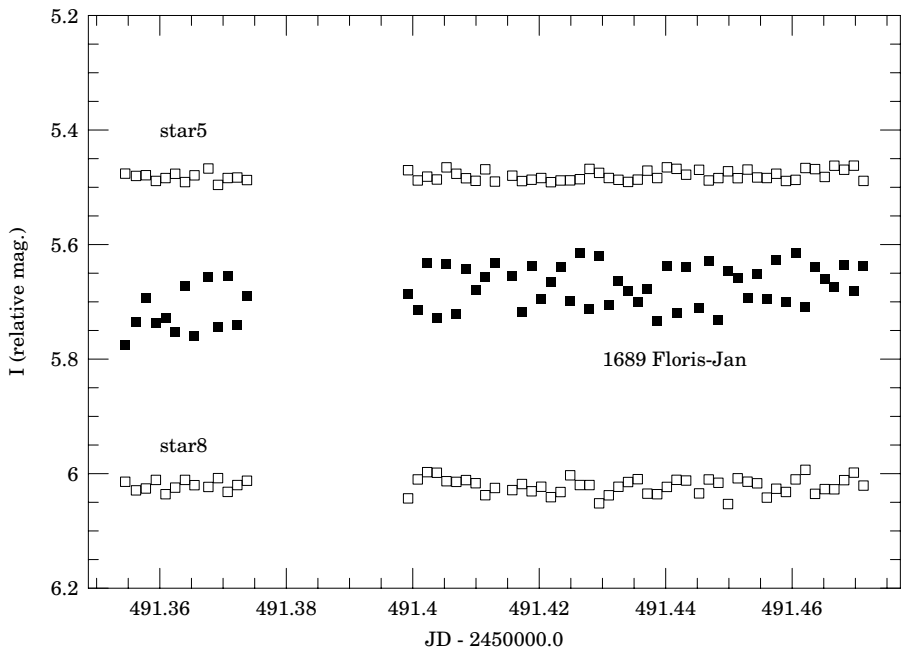}
\caption[]{Light curve of 1689 Floris-Jan compared with two light curves of a
nearby stars. Data from 1997.02.11/12}
\end{figure}
We subtracted linear trends from the data from Feb.10th, 1997 and Feb.
11th, 1997 separately. 
Figure 2 shows the AoV (Schwarzenberg-Czerny, 1991) and
Fourier-Clean (Roberts et.al., 1987)
periodograms computed on the collected data points. Peaks corresponding
to the period of 0.003461 ${\pm}$ 0.000006 [days] (4.98 ${\pm}$ 0.01 minutes)
are clearly present on both periodograms.
\begin{figure}[htbp]
\vspace{9cm}
\includegraphics{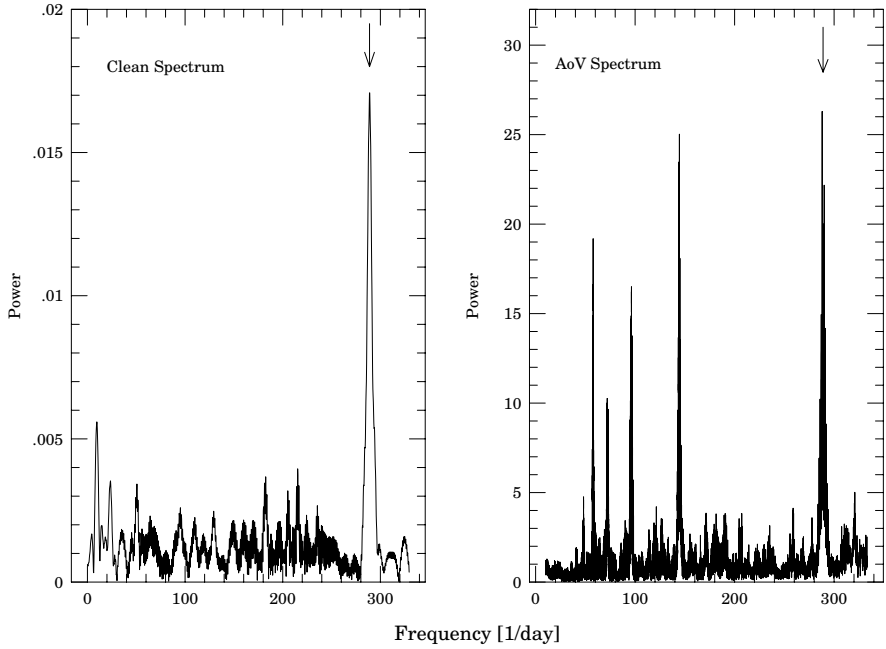}
\caption[]{1689 Floris-Jan: Clean and AoV periodograms of data points from
1997.02.10/11 and 1997.02.11/12}
\end{figure}
\begin{figure}
\vspace{9cm}
\includegraphics{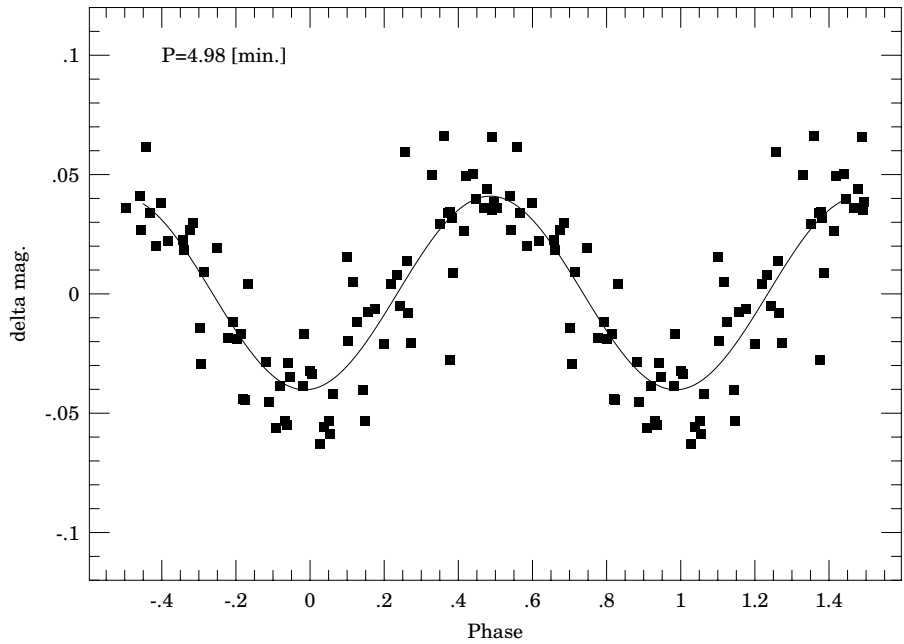}
\caption{1689 Floris-Jan: Phased light curve. Data points from
1997.02.10/11 and 1997.02.11/12 after subtraction of linear trend.}
\end{figure}
Figure 3 presents phased
light curve based on the points from the two nights. Solid line shows sine
fit to the obtained light curve. Full amplitude of the fit is 0.084 mag.
Taking into account that exposure time was 40{\%} of the observed
period we made the amplitude correction according to the following
formula:
\begin{equation}
A = A_{obs} {\cdot} \frac{\int_{0.1 \cdot \pi}^{0.9 \cdot \pi} dt} 
{\int_{0.1 \cdot \pi}^{0.9 \cdot \pi} \sin{t} dt} \approx 1.32 \cdot A_{obs}
\end{equation}
This correction yields the real amplitude of about 0.11 mag.

The observed period might be two times longer (i.e. about 10 minutes),
if we assume that it was a result of the fast rotation of the object.
Lately two independent groups claimed similar periodicity in the
light variations of an Apollo-type object 1998 KY26
(Pravec and Sarounova, 1998; Hicks and Rabinowitz, 1998).
No light oscillations of the 5 minute period were observed in
the data obtained for 1689 Floris-Jan in the night of Match 7th, 1997.
Figure 4 presents the light curve of 1689 Floris-Jan compared with the
light curves of the two nearby stars. No significant light variations are
present.
\begin{figure}
\vspace{9cm}
\includegraphics{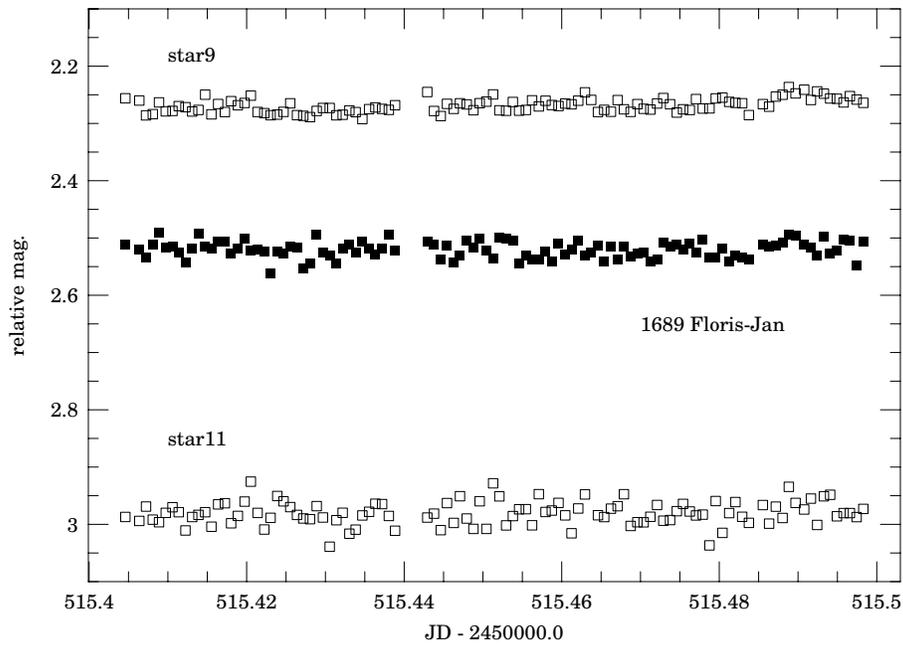}
\caption{Light curve of 1689 Floris-Jan compared with two light curves of a
nearby stars. Data from 1997.03.07/08}
\end{figure}
This suggests that ultrashort periodicity is not a permanent feature of
this object and may be present in different asteroids at different
moments of time.	
\acknowledgments{We are thankful to M. Nale{\.z}yty for his cooperation in
observations. Special thanks to Janusz Ka{\l}u{\.z}ny and Krzysztof Stanek
for their fruitful comments. We would like to thank the Warsaw
University Observatory for observing time allocation.
This work was supported from the grants KBN 2 P03D 024 09 and 2 P03D 011 12.}
\end{document}